\documentclass[fleqn,10pt]{wlscirep} 
\usepackage[utf8]{inputenc} 
\usepackage[T1]{fontenc} 
\usepackage{graphicx}
\usepackage{subcaption}
\usepackage{amsmath}
\usepackage{booktabs} 
\usepackage{siunitx}

\title{Daily and Weekly Periodicity in Large Language Model Performance and Its Implications for Research} 

\author[1,*]{Paul Tschisgale} 
\author[2]{Peter Wulff} 

\affil[1]{Leibniz Institute for Science and Mathematics Education, Kiel, Germany} 
\affil[2]{Ludwigsburg University of Education, Ludwigsburg, Germany} 
\affil[*]{Corresponding author (mail address: tschisgale@leibniz-ipn.de)} 

\begin{abstract} 
Large language models (LLMs) are increasingly used in research as both tools and objects of study. Much of this work assumes that LLM performance under fixed conditions—identical model snapshot, hyperparameters, and prompt—is time-invariant, meaning that average output quality remains stable over time; otherwise, reliability and reproducibility would be compromised.
To test the assumption of time invariance, we conducted a longitudinal study of GPT-4o’s average performance under fixed conditions. The LLM was queried to solve the same physics task ten times every three hours over approximately three months. Spectral (Fourier) analysis of the resulting time series revealed substantial periodic variability, accounting for about 20\% of total variance. The observed periodic patterns are consistent with interacting daily and weekly rhythms. These findings challenge the assumption of time invariance and carry important implications for research involving LLMs.
\end{abstract} 

\begin{document} 
\flushbottom 
\maketitle 
\thispagestyle{empty}

\section{Introduction}


Large language models (LLMs) are a specific model class in the field of generative artificial intelligence, specialized in natural language processing and generation. In the last years, there has been rapid progress in LLM development, accompanied by notable improvements in performance on a broad range of language-related tasks\cite{brownLanguageModelsAre2020,naveedComprehensiveOverviewLarge2025}.

Given these substantial advances in LLMs, researchers have shown increasing interest in examining their apparent capabilities across diverse domains, thereby treating \textit{LLMs as objects of research}. In this line of research, LLMs are often evaluated using domain-specific examinations across diverse educational levels \cite{kortemeyerCouldArtificialintelligenceAgent2023a,yeadonImpactAIPhysics2024} or established assessment instruments from educational research \cite{aldazharovaAssessingAIsProblem2024,kortemeyerMultilingualPerformanceMultimodal2025}, with their performance typically benchmarked against that of human participants. In addition, large-scale benchmark datasets have been developed in multiple domains\cite{wangSciBenchEvaluatingCollegelevel2024,fengPHYSICSBenchmarkingFoundation2025}, one prominent example being Humanity’s Last Exam, which comprises expert-level questions spanning a broad range of subjects\cite{phanHumanitysLastExam2025}.

Moreover, researchers have increasingly recognized the potential of \textit{LLMs as research tools} to support a wide range of research-related tasks. 
In particular, LLMs are increasingly adopted in qualitative research, especially for deductive coding \cite{thanUpdatingFutureCoding2025,xiaoSupportingQualitativeAnalysis2023}. In such applications, the model’s output effectively functions as a coding decision by assigning a given text segment to a category from a predefined coding scheme. Correspondingly, commercial qualitative data analysis software has begun to integrate LLM functionality. Beyond qualitative coding, LLM-based approaches are also being explored for data extraction in systematic literature reviews \cite{jansenDataExtractionGenerative2025}, reflecting a broader trend toward the integration of AI-based tools to enhance research productivity.

An implicit assumption underlying many studies that investigate LLMs as objects of research or employ them as research tools is that LLM performance is time-invariant under fixed conditions (model snapshot, hyperparameters, and prompts are fixed). By time invariance, we mean that an LLM’s average performance does not systematically depend on when it is queried. For example, if the same set of tasks is administered under identical conditions at different times during a day, systematic differences in average performance would undermine conclusions drawn from a single measurement occasion, eventually threatening the reliability, validity, and reproducibility of the research findings.

Beyond anecdotal reports from online user discussions suggesting temporal variability in LLM performance \cite{reddit_chatgpt_time_performance_2026}, emerging empirical evidence increasingly supports the existence such variability even under controlled conditions. For instance, Gupta et al.~\cite{guptaLargeLanguageModels2025} observed measurable performance changes in externally hosted LLMs across monthly testing intervals, and Tschisgale et al.~\cite{Tschisgale_6fmx-bsnl} reported significant differences in score distributions for LLM–generated physics solutions produced six weeks apart. So there is growing evidence indicating that temporal variability may persist even under fixed conditions.
Consequently, the assumption of time invariance in LLM performance should not be taken for granted and requires empirical validation.


\subsection{The Assumption of Time Invariance}
Several factors influence the output of an LLM, including the precise model snapshot (i.e., the exact internal version of the model rather than merely its general family or name), hyperparameters (such as temperature), and the user-provided prompt. Depending on the specific LLM, additional hidden system prompts typically further constrain or guide the generation process, as documented, for example, in the system description of GPT-5 \cite{openaiGPT5SystemCard2025}.

Even when all adjustable conditions are held constant, LLMs typically exhibit variability in their outputs, as response generation relies on autoregressive next-token prediction \cite{Zhang2025npjAI_ScientificMethod,Wulff.2025}. In an autoregressive framework, text is generated token by token by sampling each subsequent token from a probability distribution over the model’s vocabulary, conditioned on the input prompt and all previously generated tokens \cite{changLanguageModelBehavior2024}. This sampling procedure is directly influenced by hyperparameters such as temperature. Consequently, outputs may differ across repeated queries despite identical fixed conditions. This stochasticity of LLM-generated output is an intentional design feature, as it allows models to balance output quality and variability and to avoid degenerate or overly repetitive responses. 

In light of these considerations, an LLM can be understood as inducing a probability distribution over the space of all possible generated outputs, conditioned on the specific model snapshot, hyperparameters, the user prompt, and potential hidden system prompts. Each generated output thus constitutes a random draw from this distribution, which explains why individual outputs may differ even when the conditioning context remains unchanged.
For example, estimates of an LLM’s mathematical performance (e.g., the proportion of correctly solved problems) or its outputs in LLM-assisted qualitative deductive coding (e.g., assigned codes) may vary across repeated runs. To address this variability, researchers typically either aggregate outputs across runs (e.g., by averaging performance scores) or select the most frequent output (e.g., via majority voting) to enhance reliability and reproducibility.

However, this reasoning rests on the implicit assumption that, under fixed conditions, the probability distribution from which outputs are sampled is stable over time, that is, \textit{time-invariant}: Repeated queries to the LLM under identical conditions are assumed to correspond to independent draws from the same underlying distribution. The law of large numbers then implies that average performance scores across multiple generations, or empirical frequencies of coding decisions, converge to well-defined expected values as the number of samples increases. From this perspective, it is generally advisable to generate multiple outputs per task and to base empirical conclusions on aggregate statistics rather than on single measurements, since an individual measurement may yield an idiosyncratic performance estimate or coding decision that is not representative of the model’s typical behavior.

Violating this assumption can compromise reliability, validity, and reproducibility. Consider first sampling multiple outputs within a short time window such that the output distribution remains stable. In this case, aggregation yields a consistent estimate of the model’s average performance at that time point. However, if the same procedure is repeated later—using the identical model snapshot, hyperparameters, and prompt but under a shifted output distribution—performance estimates may systematically differ despite unchanged experimental settings, thereby undermining reproducibility.

Conversely, if sampling spans a longer time window during which the output distribution changes, increasing the number of samples does not necessarily enhance reliability. Instead of reducing random error, aggregation may average across distinct temporal regimes, producing estimates that depend on the sampling window. This weakens reliability (as repeated measurements yield systematically different results), threatens construct validity (because no single time-invariant performance parameter is being estimated), and can yield misleading conclusions about overall model performance.

\subsection{Potential Sources of Time Invariance}

For LLMs deployed on shared or centrally managed server infrastructure, time invariance may arise due to periodic patterns in server load that operate on both daily and weekly time scales. Prior work has shown that real-world usage of large-scale conversational LLM-based systems exhibits pronounced temporal regularities, with interaction volumes peaking during working hours and on weekdays, and declining during nights and weekends \cite{wangBurstGPTRealworldWorkload2025}. Similarly, seasonality has been documented for data-center workloads more generally, where server utilization follows different daily profiles on weekdays compared to weekends \cite{landreSeasonalStudyUser2024}. These usage patterns imply corresponding periodic variations in computational demand placed on LLM infrastructure. Higher demand typically leads to increased latency and lower throughput, which is undesirable from a user-experience perspective and is therefore actively managed by service providers. To maintain latency within acceptable bounds under high load, providers may employ various load-shedding or efficiency-oriented strategies, such as input compression (e.g., prompt pruning), model compression (e.g., routing requests to quantized models), or inference engine optimization (e.g., sampling from a reduced vocabulary) \cite{miaoEfficientGenerativeLarge2026, zhouSurveyEfficientInference2024}. While these measures are effective in reducing latency, they can also degrade the quality of generated outputs. Consequently, periodic variations in server load—driven by daily and weekly rhythms of user activity—could translate into periodic changes in LLM performance.

Of course, real-world LLM deployments are more complex than this simplified picture suggests. Usage patterns vary across countries and time zones, which may smooth or shift aggregate demand peaks. Moreover, providers typically operate multiple geographically distributed data centers rather than a single shared capacity, and requests may be routed dynamically across regions depending on availability, pricing, or failover policies. Additional factors such as autoscaling, heterogeneous hardware, and tiered service levels may further modulate effective capacity. These mechanisms can attenuate or redistribute temporal effects, but they do not eliminate the possibility that residual periodic performance patterns remain observable at the system level.

\subsection{Modeling of Time Variability}
\label{sec:hypothesized findings}
It is therefore possible that LLM performance varies periodically over both the day and the week. Crucially, these two rhythms are unlikely to be independent in an additive sense: the daily rhythm is likely to differ between weekdays and weekends, both in amplitude and in shape. This implies that the weekly performance pattern modulates the daily performance pattern rather than merely adding a slowly varying baseline. Accordingly, we can assume a multiplicative data-generating process—comprising a 24\,h daily component and a 7\,d weekly modulation—giving rise to an observable performance pattern that should be detectable in time-series data, for example using Fourier analysis \cite{oppenheimDiscretetimeSignalProcessing1999}.

Assuming such a multiplicative process in which a daily rhythm (24\,h) is modulated by a weekly cycle (7\,d), the resulting Fourier spectrum corresponds to the convolution of their respective spectra. Because both the daily and weekly performance rhythms are likely non-sinusoidal, they contain harmonics at integer multiples of $f_d = 1\,\mathrm{day}^{-1}$ and $f_w = 1/7\,\mathrm{day}^{-1}$, yielding interaction terms at
\[
f = k f_d \pm m f_w.
\]
These interaction terms should be observable as peaks in a Fourier spectrum obtained by Fourier-transforming the time-series data of LLM performance. Accordingly, one expects not only peaks near 24\,h and 7\,d, but also characteristic sidebands, particularly at approximately 20.6\,h and 28.4\,h for $k=m=1$.


Taken together, there is theoretical reasoning and growing empirical evidence suggesting that the performance of widely-used LLMs may not be time-invariant, with potentially serious implications for the reliability, validity, and reproducibility of research on LLM capabilities as well as studies that employ LLMs as research tools. In particular, considerations of how server demand is managed in response to usage patterns suggest that LLM performance may exhibit periodic temporal variability, potentially characterized by a 24-hour rhythm modulated by a 7-day cycle, which would challenge the assumption of time invariance. The present study empirically investigates to what extent the assumption of time invariance holds for LLM performance. To this end, a specific LLM (GPT-4o) is prompted to solve the same physics task ten times every three hours over a period of approximately three months under fixed conditions (API usage; same model snapshot, hyperparameters, and prompt). The resulting time series is analyzed using Fourier methods to examine whether the hypothesized periodic components—arising from the interaction between daily and weekly rhythms—can be detected.
In doing so, we assess whether potential performance differences across time are negligible or substantial enough to compromise the reliability, validity and reproducibility of LLM-related research findings.

\section{Results} 
A specific snapshot of GPT-4o (\texttt{gpt-4o-2024-08-06}) was queried to solve a physics task via the OpenAI API at a fixed temperature of $T=1$ using identical prompting. Ten queries were issued every three hours, beginning on August 5, 2025, at 6:00 AM and ending on October 31, 2025, at 9:00 PM (in Germany; CEST, UTC+2). Each of the overall $N = 6{,}930$ queries produced a valid response to the task, which was evaluated on a normalized scoring scale ranging from 0 (no credit) to 1 (full credit), with increments of 0.25.
\subsection{Descriptive Analysis of Variability} 

\begin{figure}[tb]
    \centering

    \begin{subfigure}[t]{0.48\textwidth}
        \centering
        \includegraphics[width=\linewidth]{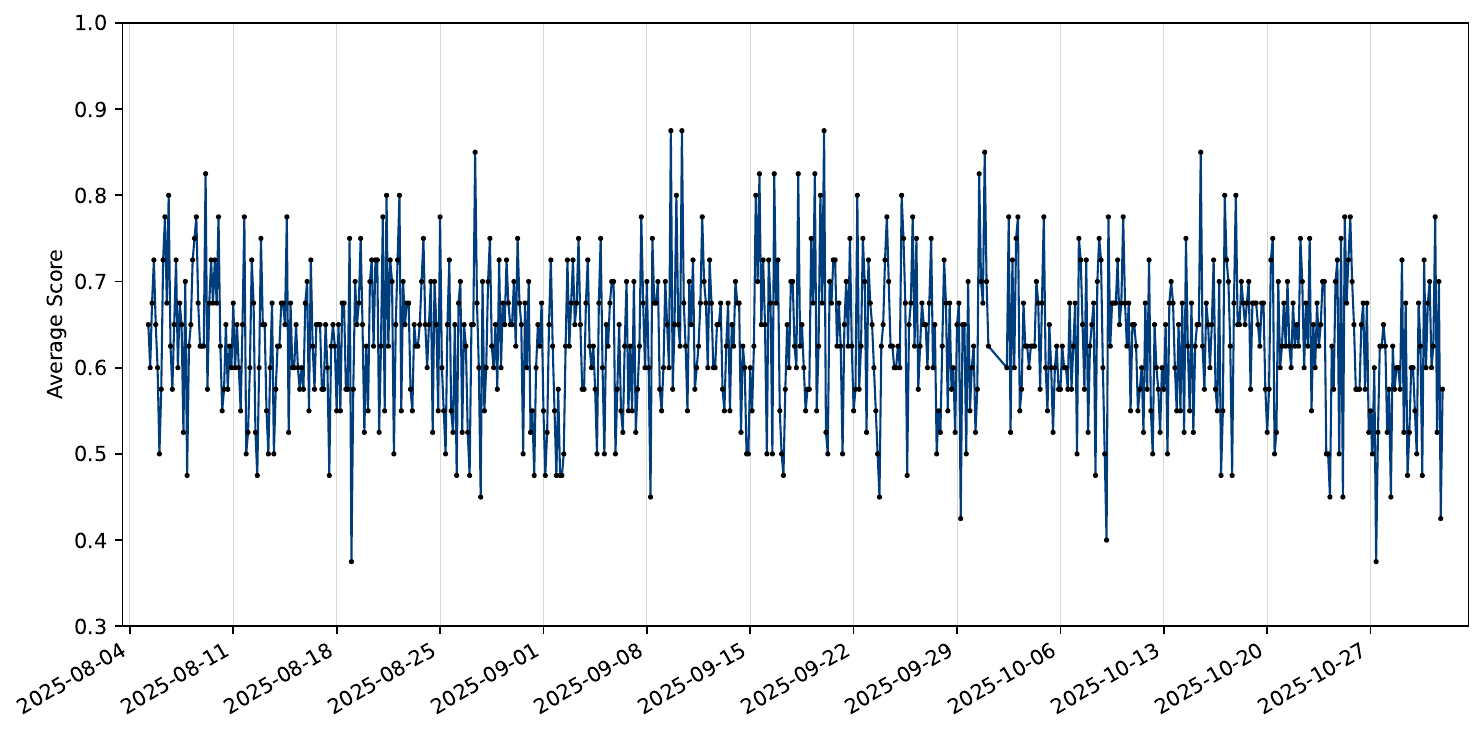}
        \caption{Full time series of LLM performance, with each dot denoting the mean score computed from 10 measurements at the respective time point. All labeled dates on the horizontal axis correspond to Mondays (CEST, UTC+2).}
        \label{fig:panel1}
    \end{subfigure}
    \hfill
    \begin{subfigure}[t]{0.48\textwidth}
        \centering
        \includegraphics[width=\linewidth]{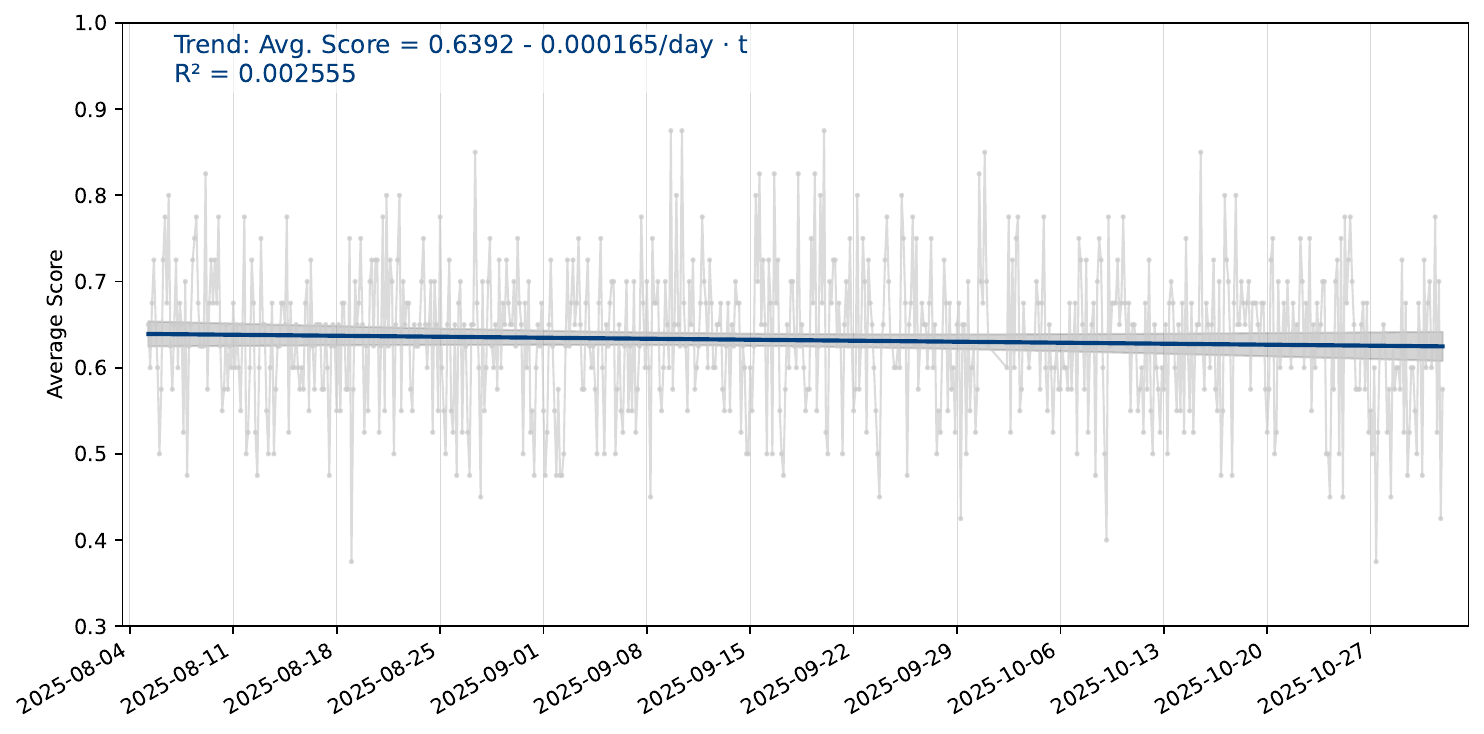}
        \caption{Linear regression line fitted to and overlaid on the full time series shown in Fig.~\ref{fig:panel1}. The shaded band indicates the 95\% confidence interval for the estimated mean trend, based on HAC standard errors.}
        \label{fig:panel2}
    \end{subfigure}

    \medskip

   \begin{subfigure}[t]{0.48\textwidth}
        \centering
        \includegraphics[width=\linewidth]{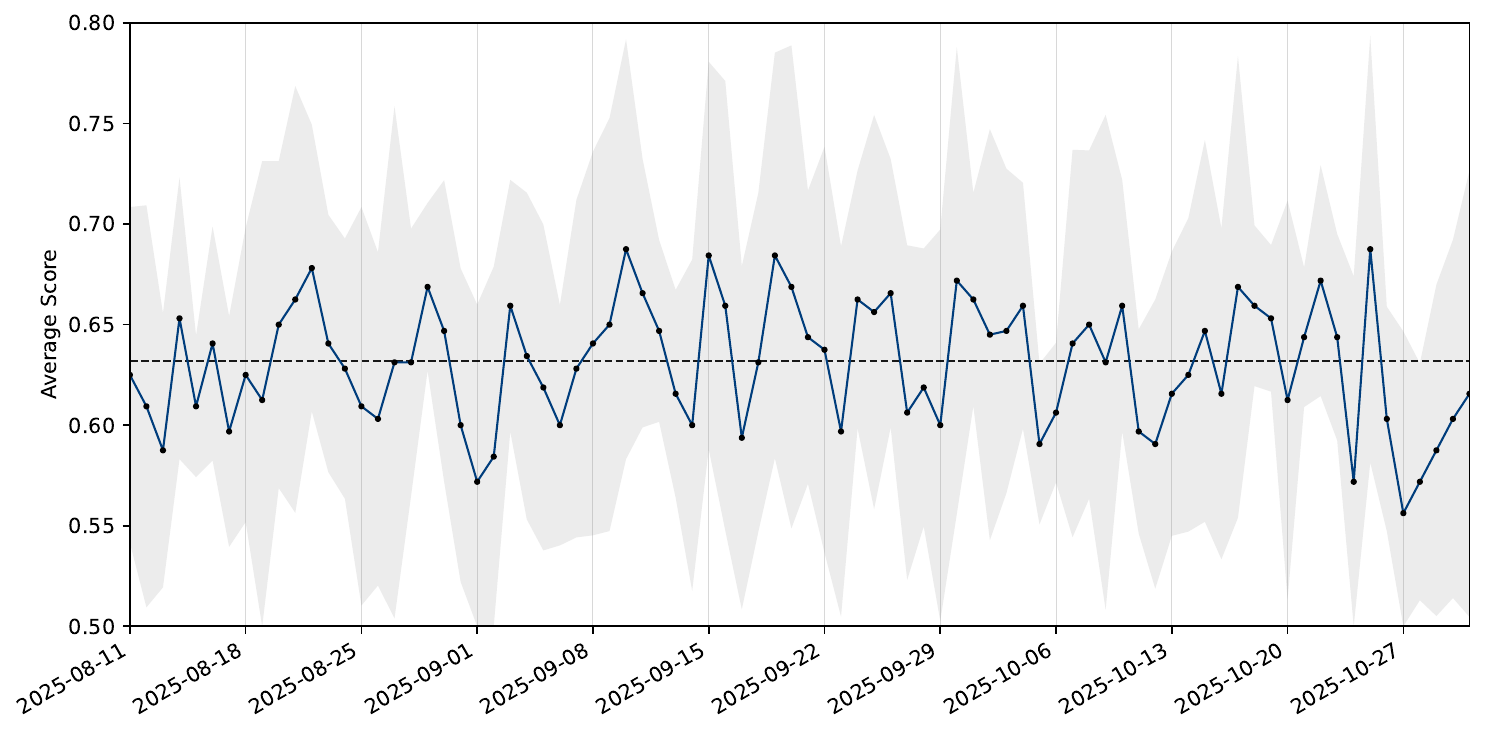}
        \caption{Daily average score over time. Dots represent daily mean scores aggregated from all measurements within each day; the shaded band indicates $\pm 1$ standard deviation. The horizontal dashed line indicates the mean of the time series.}
        \label{fig:panel3}
    \end{subfigure}
    \hfill
    \begin{subfigure}[t]{0.48\textwidth}
        \centering
        \includegraphics[width=\linewidth]{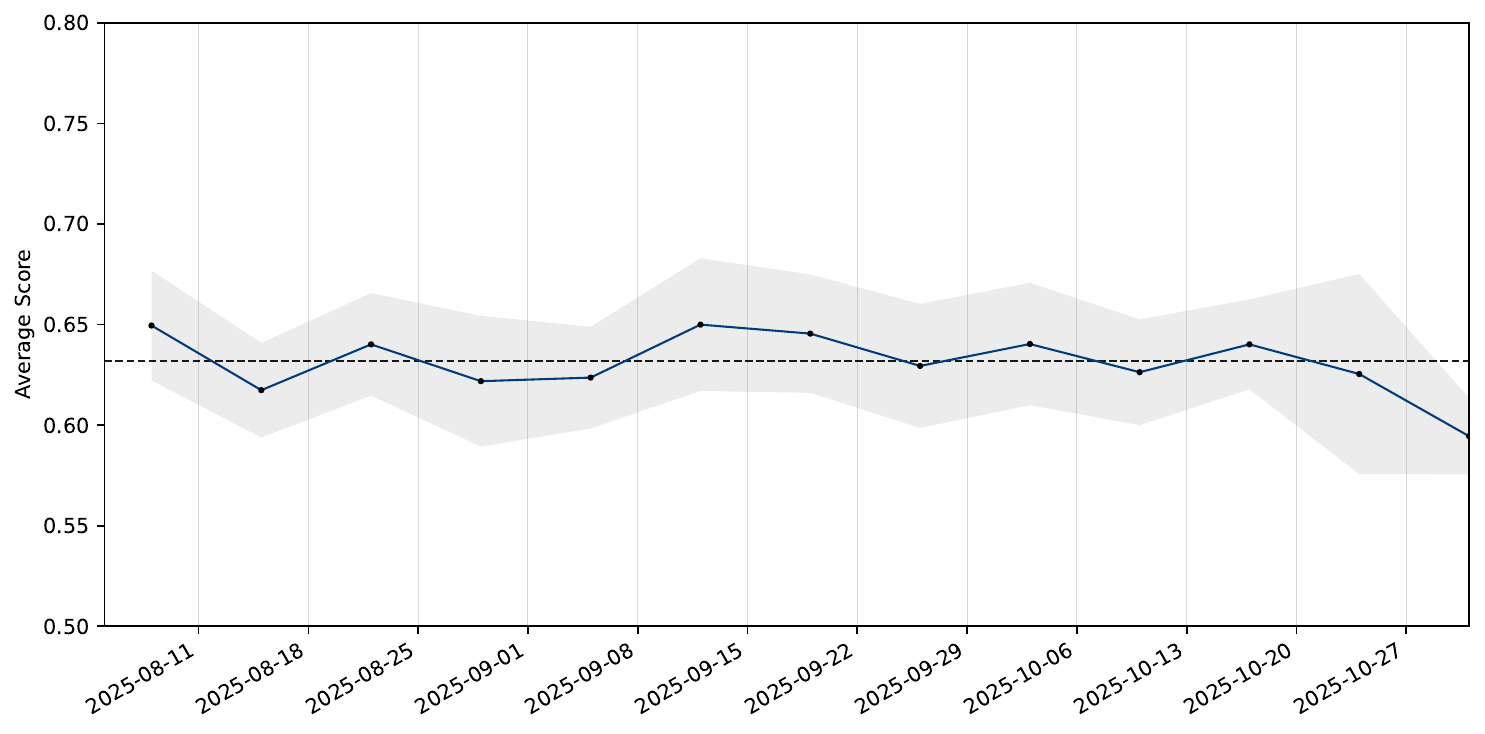}
        \caption{Weekly average score over time. Dots represent weekly mean scores aggregated from all measurements within each week from Monday to Sunday; the shaded band indicates $\pm 1$ standard deviation. The horizontal dashed line indicates the mean of the time series.}
        \label{fig:panel4}
    \end{subfigure}

    \caption{Visualization of temporal variability in the score data across different time scales.}
    \label{fig:four_panel}
\end{figure}

\begin{figure}[tb]
  \centering
  \includegraphics[width=0.7\linewidth]{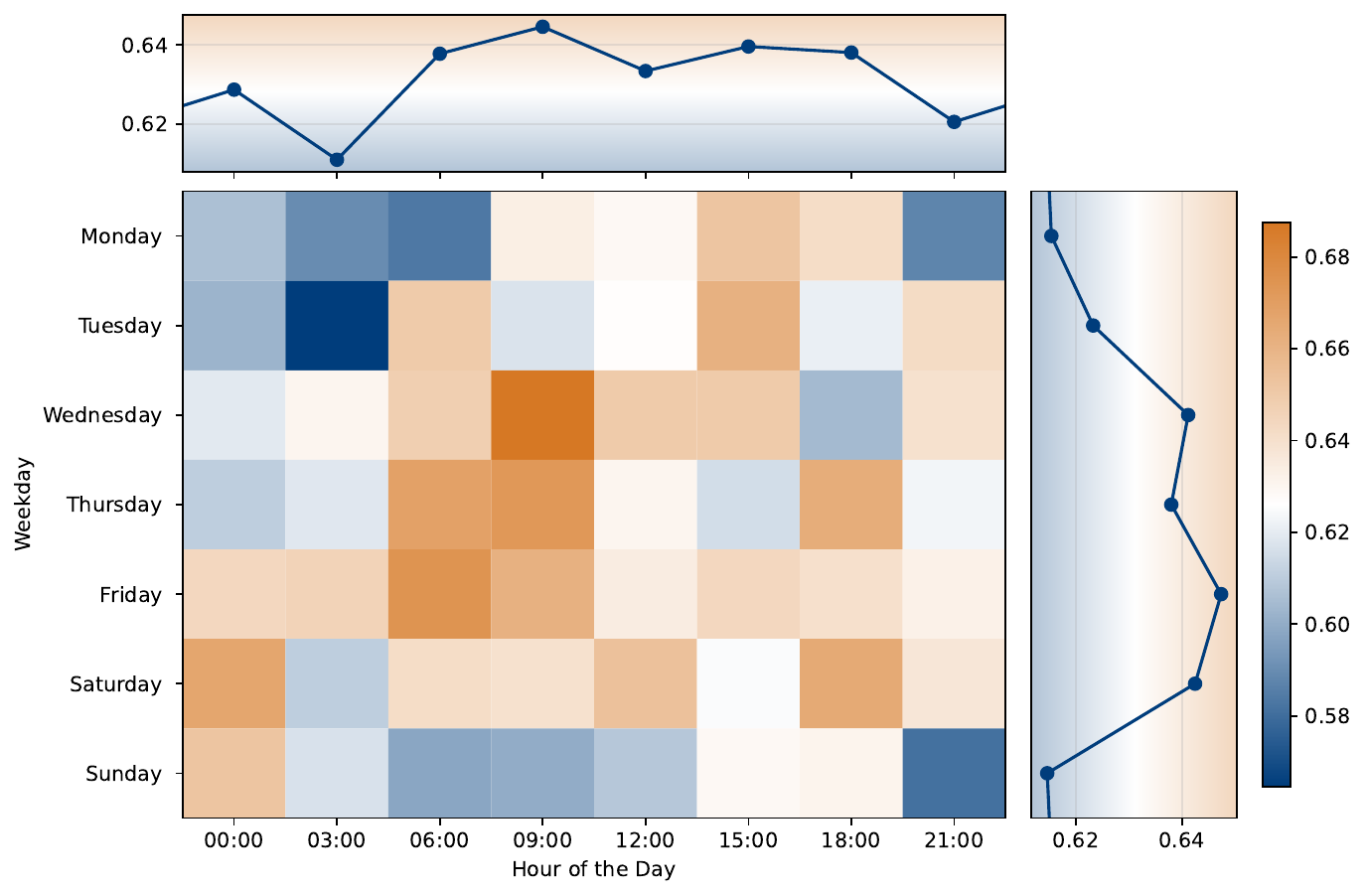}
  \caption{Heatmap of average accuracy as a function of day-of-week (rows) and hour-of-day (columns), where hour-of-day corresponds to measurement time points taken every three hours. The top panel shows the marginal average accuracy for specific hours of the day, averaged over all days of the week. The right panel shows the marginal average accuracy for each day of the week, averaged over all measured hours-of-day. All times and days are based on Central European Summer Time (CEST, UTC+2).}
  \label{fig:heatmap}
\end{figure}

Across all queries, the LLM achieved a mean accuracy of $M = 0.632$ ($SD = 0.260$) on the physics task. To reduce sampling noise beyond potential periodic variability, the ten observations collected at each time point were averaged; the resulting aggregated time series is shown in Fig.~\ref{fig:panel1}. An ordinary least squares (OLS) regression with heteroskedasticity- and autocorrelation-consistent (HAC) standard errors revealed no systematic drift in performance over time, as indicated by a non-significant drift coefficient ($t=-1.01$, $p=0.303$; see also Fig.~\ref{fig:panel2}).

To further explore the temporal structure, measurements were aggregated at the daily (Fig.~\ref{fig:panel3}) and weekly (Fig.~\ref{fig:panel4}) levels. Daily averages exhibited noticeable variability across days of the week, whereas weekly averages deviated only marginally from the overall mean, suggesting no clear periodic patterns beyond the weekly time scale.
To examine daily and weekly variation jointly, we aggregated the data for each combination of day-of-week and hour-of-day. The resulting averages are shown in the heatmap in Fig.~\ref{fig:heatmap}, together with the corresponding marginal distributions. Visual inspection indicates that time-of-day performance patterns vary across day-of-week, suggesting an interaction between daily and weekly temporal patterns, such that daily performance rhythms appear to be modulated by weekday.



\subsection{Fourier Analysis of Periodic Variability}
\begin{figure}[!tb]
  \centering \includegraphics[width=0.8\linewidth]{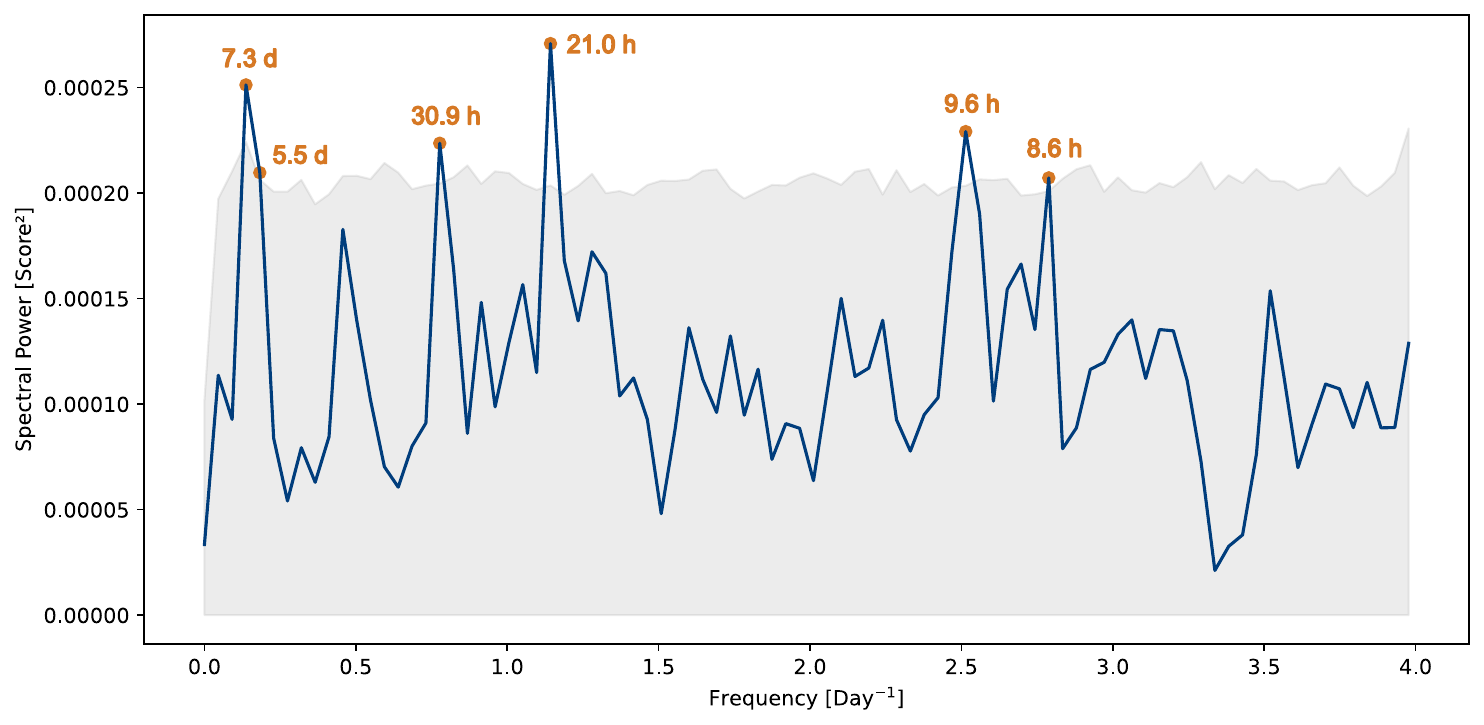}
  \caption{Power spectrum estimated via fast Fourier transformation using Welch’s method and Hann-windowing. The grey shaded band indicates the 95\% permutation-based significance threshold; labeled spectral peaks exceeding this band are considered statistically significant.}
  \label{fig:welch_spectrum}
\end{figure}

We conducted a Fourier analysis using the fast Fourier transform in combination with Welch’s method to identify dominant periodic components in the time-series data \cite{cochranWhatFastFourier1967, welchUseFastFourier1967}.
Results of the Fourier analysis in the form of a power spectrum are shown in Fig.~\ref{fig:welch_spectrum}. The spectrum exhibits several distinct peaks (see Table~\ref{tab:fourier_peaks}) that exceed the permutation-based significance threshold, indicating the presence of statistically significant periodic components.

Two neighboring significant peaks at approximately 5.5\,d and 7.3\,d appear as a single broadened spectral peak, likely attributable to limited frequency resolution and spectral leakage, and are consistent with a weekly periodic component.
Further significant peaks are observed at periods of approximately 30.9\,h and 21.0\,h, while no dominant peak appears at exactly 24\,h. Instead, the peaks flank the 24\,h period, consistent with sidebands produced by a daily rhythm modulated by a weekly cycle.
Formally, if a daily component with frequency $f_d = 1\,\mathrm{day}^{-1}$ is multiplicatively modulated by a weekly component with frequency $f_w = 1/7\,\mathrm{day}^{-1}$, spectral peaks are expected at $f = f_d \pm f_w$, corresponding to periods of approximately $28.0\,\mathrm{h}$ and $21.0\,\mathrm{h}$. The observed $21.0\,\mathrm{h}$ peak closely matches the predicted upper sideband, whereas the $30.9\,\mathrm{h}$ peak deviates from the lower sideband, likely due to limited frequency resolution and spectral leakage. The absence of a sharp $24\,\mathrm{h}$ peak suggests that the daily component does not act as an independent oscillation but varies systematically across the week (as can be seen in Fig.~\ref{fig:heatmap}). Consequently, daily-scale variability manifests primarily through sidebands around $24\,\mathrm{h}$ rather than as a distinct $24\,\mathrm{h}$ spectral peak.

Finally, significant sub-daily peaks are observed at approximately 9.6\,h and 8.6\,h. These components likely reflect harmonics of a non-sinusoidal daily rhythm (see Fig.~\ref{fig:heatmap}). In particular, the second harmonic ($2 f_d$, 12\,h) and third harmonic ($3 f_d$, 8\,h), when modulated by the weekly frequency $f_w = 1/7\,\mathrm{day}^{-1}$, generate sidebands at $k f_d \pm f_w$ for $k=2,3$, corresponding to periods near 10.6/9.2\,h and 7.6/8.4\,h, respectively. Given finite frequency resolution and spectral leakage, power from these sidebands may manifest near the observed 9.6\,h and 8.6\,h peaks. More generally, modulation can render individual sidebands detectable even when the underlying harmonic itself remains below significance.

\begin{table}[tb]
\centering
\caption{Information on the detected significant peaks in the power spectrum shown in Fig.~\ref{fig:welch_spectrum}. Phase was estimated separately by least-squares sinusoidal fitting at each frequency (cosine reference).}
\label{tab:fourier_peaks}
\begin{tabular}{ccccc}
\toprule
\shortstack{\textbf{Period} \\ \textbf{Length}} &
\shortstack{\textbf{Frequency} \\ \textnormal{[$\text{day}^{-1}$]}} &
\shortstack{\textbf{Spectral Power} \\ \textnormal{[$\text{score}^2$]}} &
\shortstack{\textbf{Amplitude} \\ \textnormal{[score]}} & 
\shortstack{\textbf{Phase} \\ \textnormal{[degree]}} \\

\midrule
\SI{7.3}{\day}             &          0.137        &         0.000251         &       0.0159       &   35.6   \\
\SI{5.5}{\day}             &          0.183        &           0.000210      &        0.0145       & -163.6   \\
\SI{30.9}{\hour}             &        0.777          &      0.000224            &         0.0150     &  9.8    \\
\SI{21.0}{\hour}           &        1.14          &         0.000271         &          0.0165    & 106.7    \\
\SI{9.6}{\hour}              &           2.51       &        0.000229          &        0.0151      & -50.6    \\
\SI{8.6}{\hour}              &        2.79          &         0.000207         &         0.0144      & 4.1   \\
\bottomrule
\end{tabular}
\end{table}

The proportion of total variability in the time series attributable to the identified significant periodic components is $20.3\%$. Moreover, the aggregate contribution of the identified periodic components in the time domain corresponds to a peak-to-peak variation of $0.139$ units in the performance score (range 0–1), indicating that periodic structure alone accounts for performance fluctuations of approximately 14\% of the full scale.

\section{Discussion} 

Our findings show that the tested LLM exhibits periodic variability in its performance on the probed physics task, characterized by an interaction between daily- and weekly-scale dynamics. 
Specifically, we found that in our context approximately 20\% of the total variance in average performance is attributable to periodic components, and that the periodic structure alone induces peak-to-peak performance variability of about 14\% of the full performance scale. These magnitudes indicate that these periodic patterns are not merely statistical artifacts but represent substantively meaningful variability in LLM behavior over time.

The following discussion focuses on the implications of our findings for LLM-based research. A detailed analysis of the temporal pattern of performance peaks and lows (see Fig. 3), including their potential relation to global usage cycles and server load, is provided in the Supplementary Information.

Our findings have important implications for the interpretation of performance estimates reported in the existing literature on widely-used LLMs' capabilities. In particular, studies based on data collection restricted to a narrow temporal window may yield biased estimates if that window coincides with periods of systematically higher or lower LLM performance. In such cases, reported performance levels may overstate or understate the true long-run average, respectively. 

When LLMs are used as research tools—for example, in tasks such as deductive qualitative coding or automated annotation—similar considerations apply. In these settings, temporal variability in model performance can introduce systematic biases into research outputs if data collection is confined to limited or unrepresentative time windows. As a result, coding decisions, category assignments, or derived measures may reflect systematic variability in model behavior rather than stable properties of the underlying data or coding scheme. This risk is particularly salient when LLM outputs are treated as fixed or objective inputs to subsequent analyses, as the temporal variability may propagate downstream and affect substantive research conclusions.

Taken together, our findings highlight a previously underappreciated source of uncertainty: sampling decisions in time. Temporal sampling choices can meaningfully affect final performance estimates, yet this source of variability has rarely been explicitly considered or quantified in prior work. As a consequence, existing estimates of LLM performance should be interpreted with greater uncertainty than is typically acknowledged.

Sampling confined to short or irregular time windows may capture transient performance levels rather than the model’s typical behavior, thereby biasing aggregate estimates of expected or average performance. To enhance validity and reproducibility in research with especially sensitive LLM usage, data collection should therefore span one full week—the longest periodicity observed—or multiples thereof. However, this has to be counterweight with environmental considerations, such as energy expenditure of LLM usage. Then, measurements should be evenly spaced across this interval, with at least daily sampling and, ideally, hourly resolution to adequately capture higher-frequency temporal structure and to obtain unbiased estimates of true expected LLM performance. In addition, multiple repetitions per time point are necessary to mitigate sampling noise arising from the stochastic nature of LLM outputs. In our data, approximately 80\% of the variability remained unexplained by temporal periodicity, even with 10 repetitions per time point, indicating that larger sample sizes may often be required. Although such designs yield a substantial number of LLM queries, automated evaluation renders them feasible in typical benchmark settings. Finally, researchers should report measures of variability and, where possible, propagate the associated uncertainty in downstream analyses.

The present study has some limitations that also point to clear avenues for future research.
First, the temporal sampling resolution was limited: Measurements were taken every three hours, restricting the detection of higher-frequency components and increasing the risk of aliasing. Moreover, only ten queries were issued per time point, resulting in still high sampling variability. Future studies should sample at least hourly and increase measurements per time point to obtain a more precise estimate of the underlying performance pattern.
Second, our analysis was limited to a single task—a multiple-choice physics problem. To evaluate the generality of the observed periodicities, future work should apply the same approach across a broader range of domains and task types.
Third, while we interpret the observed daily and weekly patterns in terms of infrastructure-level load management and efficiency-oriented inference strategies, this does not account for a causal explanation of the observed temporal variability in performance. A crucial next step is therefore to compare externally hosted LLMs from different providers, which may implement different load-balancing and optimization schemes, as well as to contrast them with locally hosted models that run without shared server-side load. If the periodic structure is driven primarily by server load and load-shedding mechanisms, it should be attenuated or absent in locally deployed models and should differ systematically across external services. More broadly, respective findings may motivate a shift toward offline, locally hosted models where feasible, as such deployments offer greater control over execution conditions and can help ensure the reproducibility and temporal stability of LLM-based analyses.

Finally, periodic temporal variability is not unique to LLMs but is well established in human cognition and judgment. Human performance similarly varies periodically across the day and week as a function of circadian rhythms, time awake, and cumulative fatigue, affecting attention, executive control, and decision consistency, even in expert populations \cite{hannonInfluencesDayWeek2016, schmidtTimeThinkCircadian2007}. Empirical evidence further demonstrates time-dependent effects in applied settings, such as judicial decision-making \cite{Danziger.2011}, as well as persistent variability in qualitative coding despite formalized procedures \cite{OConnor.2020}. A key distinction, however, concerns metacognitive monitoring: Whereas humans can in principle reflect on and regulate their cognitive states, albeit imperfectly, LLMs lack any capability for self-monitoring  fluctuations in their own performance, implying that temporally variability—once present—cannot be internally detected or compensated for by the system itself. In this fundamental respect, humans retain a decisive advantage.
\section{Methods} 

\subsection{Task Selection and Scoring}
To measure performance of an LLM over time, we selected a physics task taken from the German Physics Olympiad at an intermediate difficulty level. Pilot tests revealed that this particular task was neither trivial (such that the selected LLM would almost always answer correctly) nor excessively difficult (such that the LLM would almost always answer incorrectly). The task is formulated as a multiple-choice question, meaning that at least one of the listed options is correct. The task—translated from German to English by the authors—reads:

\begin{quote}
\small
    A single battery can power an incandescent bulb for a time $t$. For simplicity, assume that the bulb shines with constant brightness until the battery is depleted, and that the resistance of the bulb is constant. Which of the following statements are correct if two of these batteries are used to operate two of the bulbs?
\begin{enumerate}
\setlength{\itemsep}{0pt}
    \item[A. ] If the batteries are connected in series and the bulbs are connected in series, the bulbs can be operated for approximately a time $t/4$.
    \item[B. ] If the batteries are connected in series and the bulbs are connected in parallel, the bulbs can be operated for approximately a time $t/2$.
    \item[C. ] If the batteries are connected in parallel and the bulbs are connected in series, the bulbs can be operated for approximately a time $2t$.
    \item[D. ] If the batteries are connected in parallel and the bulbs are also connected in parallel, the bulbs can be operated for approximately a time $t$.
\end{enumerate}
\end{quote}
Applying basic electrical circuit theory and energy considerations shows that only answer option D is correct.

During pilot tests, we observed that the LLM appeared to evaluate each answer option independently in terms of its plausibility. Accordingly, we adopted an option-wise scoring scheme in which each answer option was evaluated independently. For each option, a score of 0.25 points was awarded for a correct decision—either selecting the option if it is correct or not selecting it if it is incorrect. This procedure yields total scores in the set 
$\{0,0.25,0.5,0.75,1\}$. To illustrate, consider the LLM response “B, D”. The response receives 0.25 points for correctly not selecting option A, 0 points for incorrectly selecting option B, 0.25 points for correctly not selecting option C, and 0.25 points for correctly selecting option D, resulting in a total score of 0.5.

\subsection{Data Generation Under Fixed Conditions}
A specific snapshot of GPT-4o (\texttt{gpt-4o-2024-08-06}) was queried via the OpenAI API (application programming interface) at a fixed temperature setting ($T=1$) using identical prompting (including system prompt, user prompt, and the introduced multiple-choice physics problem). The API enables programmatic access to OpenAI models from software environments such as Python, allowing recurring queries (e.g., every three hours) to be automated over extended periods of time. 
This way, ten queries were issued every three hours, beginning on August 5, 2025, at 6:00 AM and ending on October 31, 2025, at 9:00 PM (in Germany; CEST, UTC+2). 

To enable automated scoring of model responses, we enforced a structured output format in which the model returned a JSON object containing (a) the complete solution path and (b) the selected multiple-choice option(s). Automated evaluation was performed solely on the basis of the selected option(s). The multiple-choice task text was provided as the user prompt. The system prompt—translated from German to English by the authors—reads as follows: “You are a physics expert. Solve the problem. Return both the detailed solution path and the final answer in JSON format.” This way, $N=6{,}930$ solutions to the problem at hand were generated and automatically scored.

\subsection{Data Analysis}
All analyses were performed using Python 3.12 \cite{python312}.
\subsubsection{Analysis of Temporal Drift}

To assess whether a systematic performance drift was present in the time-series data, we fitted an ordinary least squares (OLS) regression to the observed values. Statistical inference on the regression coefficients, in particular on the linear drift term, was conducted using Newey–West heteroskedasticity- and autocorrelation-consistent (HAC) standard errors \cite{neweySimplePositiveSemidefinite1987}. This approach provides consistent covariance estimates in the presence of both heteroskedasticity and autocorrelation in the regression residuals, which are common in time-series data. The truncation lag (bandwidth) was selected to cover the longest expected temporal dependence in the data (7 days). Additionally, all measurements were averaged at the daily level and at the weekly level to visually examine potential temporal trends and patterns.

\subsubsection{Missing Data Handling}

Due to unforeseen technical difficulties, data generation was interrupted for a short period. Specifically, no API queries were recorded between October 1, 2025, 06:00 and October 2, 2025, 06:00 CEST, resulting in nine missing time points, each comprising 10 repeated measurements (which can also be seen by inspecting Fig.~\ref{fig:panel1}). To handle these missing observations, we applied mean-value imputation using the overall mean of the complete time series, because the missing interval comprises only a small fraction of the total observation window and is therefore unlikely to materially influence estimates of trends, variability, or periodicity, regardless of the specific imputation method used.

\subsubsection{Fourier Analysis}

\begin{figure}[tbp]
    \centering

    \begin{subfigure}[t]{0.45\textwidth}
        \centering
        \includegraphics[width=\linewidth]{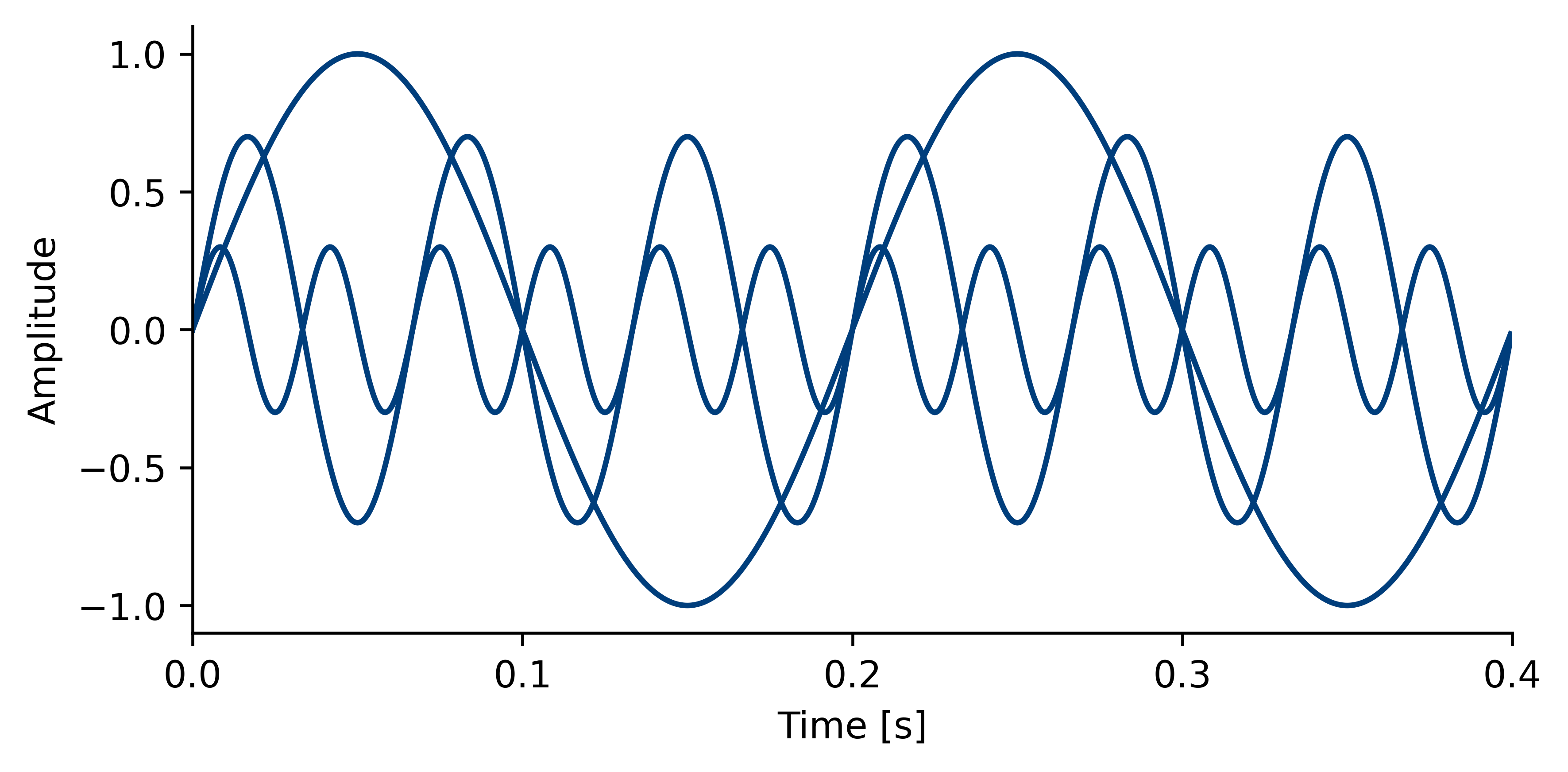}
        \caption{Three sinusoidal signals of the form $x_i(t)=A_i \sin(2\pi f_i t)$ with amplitudes $A_1=1.0$, $A_2=0.7$, and $A_3=0.3$, and frequencies $f_1=5\,\mathrm{s}^{-1}$, $f_2=15\,\mathrm{s}^{-1}$, and $f_3=30\,\mathrm{s}^{-1}$.}
        \label{fig:top}
    \end{subfigure}

    \vspace{1em}

    \begin{subfigure}[t]{0.45\textwidth}
        \includegraphics[width=\linewidth]{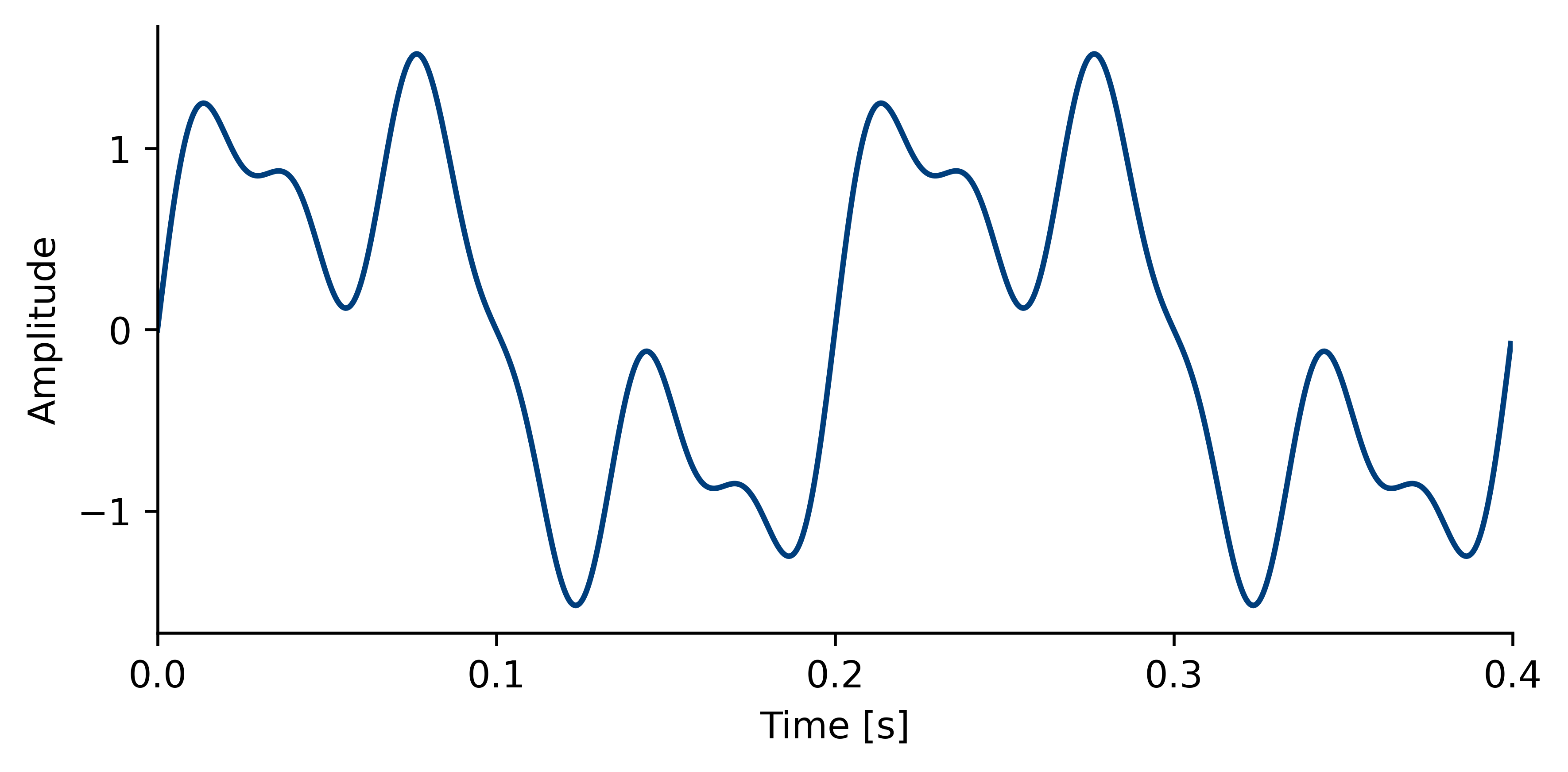}
        \caption{Combined signal $x(t)=x_1(t)+x_2(t) +x_3(t)$ in the time domain.}
        \label{fig:bl1}
    \end{subfigure}
    \hfill
    \begin{subfigure}[t]{0.45\textwidth}
        \centering
        \includegraphics[width=\linewidth]{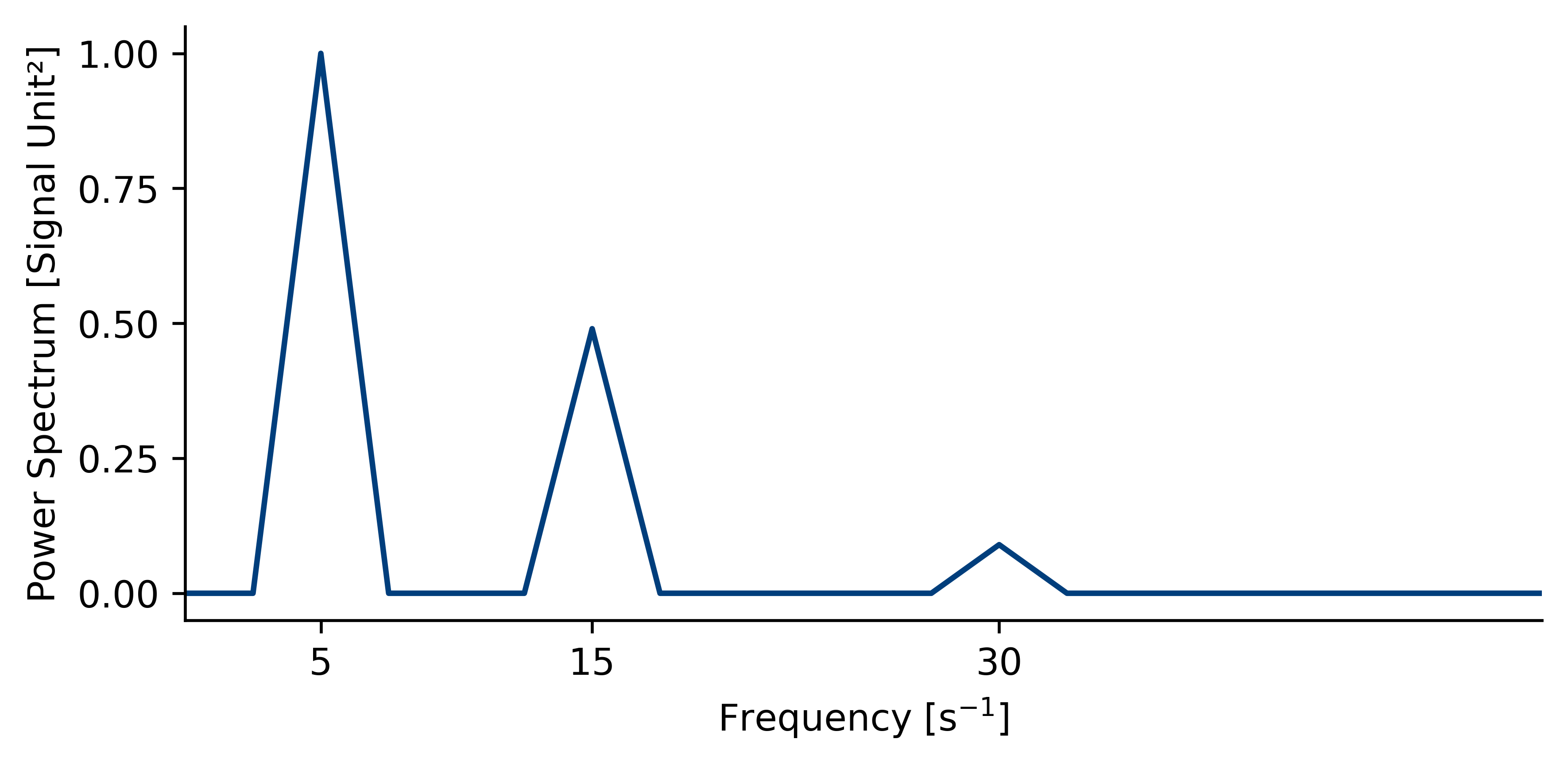}
        \caption{Power spectrum obtained from the Fourier transform of the combined signal $x(t)=x_1(t)+x_2(t)+x_3(t)$ depicted in Fig.~\ref{fig:bl1}, showing three dominant frequency components at $5\,\mathrm{s}^{-1}$, $15\,\mathrm{s}^{-1}$, and $30\,\mathrm{s}^{-1}$. Peak power equals the squared time-domain amplitude, i.e., $1.0=1.0^2$, $0.49=0.7^2$, and $0.09=0.3^2$.}
        \label{fig:br1}
    \end{subfigure}

    \vspace{0.5em}

    \begin{subfigure}[!tbp]{0.45\textwidth}
        \centering
        \includegraphics[width=\linewidth]{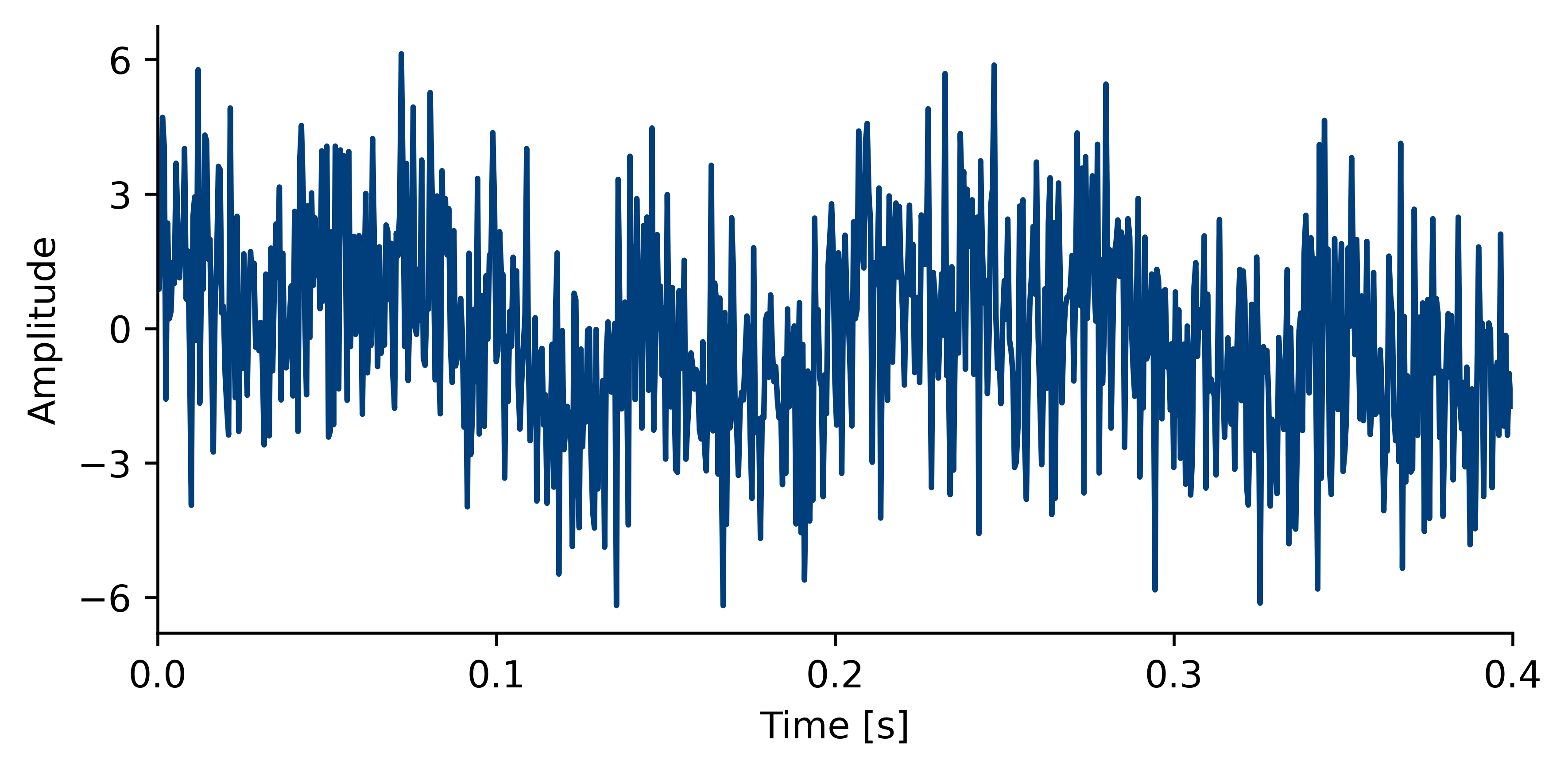}
        \caption{Noisy combined signal $\tilde{x}(t)=x(t)+\varepsilon(t)$ in the time domain, where $\varepsilon(t)\sim\mathcal{N}(0,\sigma^2)$ with $\sigma=2$ denotes additive white Gaussian noise.}

        \label{fig:bl2}
    \end{subfigure}
    \hfill
    \begin{subfigure}[t]{0.45\textwidth}
        \centering
        \includegraphics[width=\linewidth]{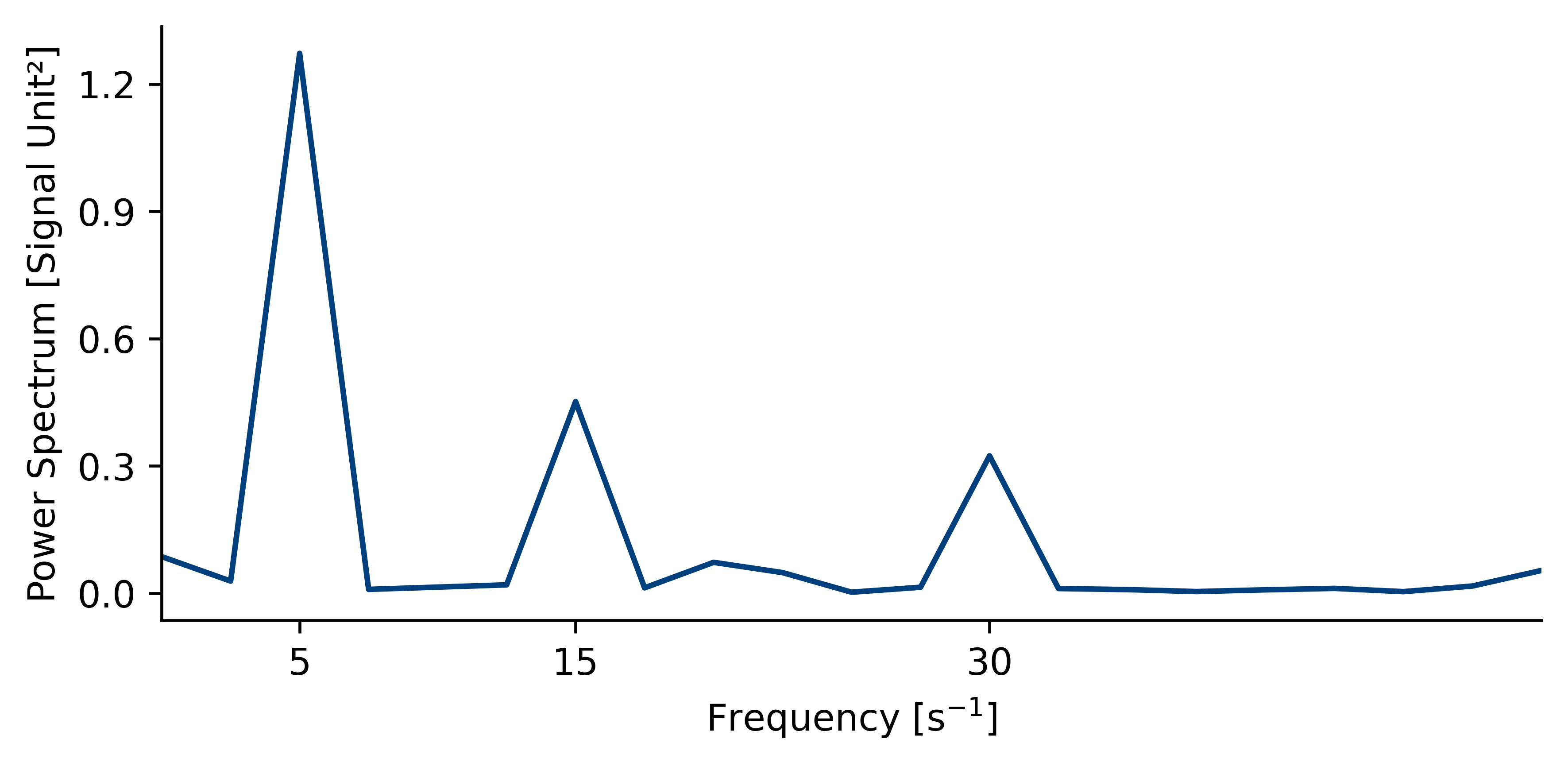}
        \caption{Power spectrum obtained from the Fourier transform of the noisy combined signal $\tilde{x}(t)=x(t)+\varepsilon(t)$ depicted in Fig.~\ref{fig:bl2}. Compared to Fig.~\ref{fig:br1}, the noise introduces a broadband background and leads to less sharply defined spectral peaks with time-domain amplitudes that deviate from the actual amplitudes of the underlying sinusoidal components.}
        \label{fig:br2}
    \end{subfigure}

    \caption{Conceptual illustration of how a spectral (Fourier) analysis operates for analyzing time-series data.}
    \label{fig:five_panel}
\end{figure}

To examine periodic structure in the time series, we performed a Fourier analysis using the fast Fourier transform \cite{cochranWhatFastFourier1967}. The basic principle of Fourier analysis is illustrated in Fig.~\ref{fig:five_panel}. Consider several periodic components, represented as sinusoidal signals with different frequencies and amplitudes (Fig.~\ref{fig:top}). When these components are additively combined, the resulting signal remains periodic but exhibits a more complex waveform, from which the individual constituent frequencies are not readily identifiable by visual inspection alone (Fig.~\ref{fig:bl1}). Fourier analysis addresses this problem by decomposing the observed signal into a weighted sum of sinusoidal basis functions, thereby representing the signal in the frequency domain (Fig.~\ref{fig:br1}). The Fourier transform thus quantifies the contribution of each frequency component to the overall signal. In practical applications, time series data are typically contaminated by noise, which obscures the underlying periodic structure (Fig.~\ref{fig:bl2}). Nevertheless, Fourier analysis remains capable of identifying dominant periodic components, albeit with reduced precision and increased uncertainty in the presence of noise (Fig.~\ref{fig:br2}).

In our case, the input data comprised an evenly sampled time series with a fixed temporal resolution of 3 hours, corresponding to a sampling frequency of $f_s = 8$ samples per day. The series spanned an observation period of approximately 87.75 days and, given the fixed sampling interval, consisted of $N = 702$ time points. At each time point, ten individual measurements were available, yielding a total of $7{,}020$ measurement-level data points; missing values at this level were imputed prior to aggregation. The maximum detectable signal frequency (Nyquist frequency) was therefore $f_{\mathrm{Nyq}} = f_s / 2 = 4$ cycles per day, corresponding to a minimum resolvable period of 6 hours.

More specifically, power spectra were estimated using Welch’s method \cite{welchUseFastFourier1967} as implemented in \texttt{scipy.signal.welch} \cite{2020SciPy-NMeth}, because this method provides a robust and low-variance estimate of the power spectrum for finite and potentially noisy time series by averaging spectral estimates across overlapping, windowed segments. This approach reduces sensitivity to random fluctuations and spectral leakage compared to a single discrete Fourier transform.

According to Welch's method, the time series was segmented into overlapping windows of fixed length ($n_\text{perseg}=\lfloor N/4\rfloor=175$), with adjacent windows overlapping by 50\%. A Hann window was applied to each segment to reduce spectral leakage \cite{prabhuWindowFunctionsTheir2018}, and the resulting segment-wise spectra were averaged to obtain the final spectral estimate. Spectral results are reported in terms of spectral power, corresponding to window-normalized squared Fourier amplitudes. Given a sampling frequency of 8 samples per day and a segment length of 175 samples, the resulting frequency grid had a spacing of approximately 0.05 $\text{day}^{-1}$, which defines the effective frequency resolution of the analysis.

To assess the statistical significance of spectral peaks, we employed a nonparametric permutation procedure \cite{odellPermutationTestPeriodicities1975, ptitsynPermutationTestPeriodicity2006}. Specifically, the observed time series was randomly permuted $n=1000$ times to generate surrogate datasets that destroyed temporal dependencies while retaining the distribution of observed values. For each surrogate, the Welch spectrum was recomputed using identical parameters. An empirical significance threshold was then constructed for each frequency bin as the 95th percentile of the bootstrap distribution, corresponding to a one-sided test with $\alpha=0.05$. Frequencies at which the observed spectrum exceeded this threshold are considered statistically significant.

Moreover, the proportion of performance variability explained by periodic components in the original time series was quantified. To this end, spectral power \(P(f)\) was summed across all frequency bins corresponding to statistically significant peak frequencies \(f^\ast\). Because the total variance of a time series equals the integral of its power spectrum \cite{oppenheimDiscretetimeSignalProcessing1999}, normalizing the sum by the total variance yields the fraction of variance attributable to the identified significant periodic patterns. Specifically, we computed
\begin{equation*}
    \text{explained variability}
    =
    \frac{\sum_{f^\ast} P(f)}{\text{Var}(X_t)} \, ,
\end{equation*}
where $\text{Var}(X_t)$ represents the variance of the respective time series. 

In addition to this measure of explained variance, we quantified the time-domain magnitude of variability attributable to the combined periodic components in order to assess how strongly their joint contribution explains fluctuations in the observed performance scores. For this purpose, phase information—unavailable from the Welch power spectrum—was estimated by least-squares fitting of a cosine–sine model to the full time series at each significant frequency. Using these phase estimates together with the corresponding amplitudes, a composite signal was reconstructed by summing all identified periodic components. The reconstructed signal was evaluated over a sufficiently long temporal interval to approximate its asymptotic range.

\section*{Acknowledgements} 
The authors are grateful to their colleagues for valuable discussions regarding reproducibility standards in AI-based research. They particularly thank Dietmar Block for insightful and constructive discussions on spectral analysis.

ChatGPT (OpenAI) was used for language editing to improve clarity and style. The authors reviewed and take full responsibility for the final content of the manuscript. 

\section*{Author contributions statement} 
PT: study conception and design, data collection, analysis, interpretation, and manuscript preparation. PW: study conception and design, data collection, and manuscript preparation. 

\section*{Data Availability} 
The data along with the analysis code are available in an OSF repository: \hyperlink{https://doi.org/10.17605/OSF.IO/PFYQ4}{https://doi.org/10.17605/OSF.IO/PFYQ4}. The repository further contains a Binder link that launches a ready-to-use computational environment for reproducing all analyses without additional setup.

\section*{Additional information} 
\textbf{Competing interests:} The authors declare no competing interests.\vspace{1\baselineskip}

\noindent\textbf{Funding:} This work was supported by the Klaus-Tschira-Stiftung (project WasP) under Grant No. 00.001.2023.

\bibliography{sample}

\end{document}